\documentclass[english,prb,twocolumn,superscriptaddress]{revtex4}
\usepackage{graphicx}
\begin{document}

\title{Existence of an independent phonon bath in a quantum device}
\author{L. M. A. Pascal}
\author{A. Fay}
\author{C. B. Winkelmann}
\author{H. Courtois}
\affiliation{Institut N\'eel, CNRS, Universit\'e Joseph Fourier and Grenoble INP, 25 avenue des Martyrs, Grenoble, France}

\begin{abstract}
At low temperatures, the thermal wavelength of acoustic phonons in a metallic thin film on a substrate can widely exceed the film thickness. It is thus generally believed that a mesoscopic device operating at low temperature does not carry an individual phonon population. In this work, we provide direct experimental evidence for the thermal decoupling of phonons in a mesoscopic quantum device from its substrate phonon heat bath at a sub-Kelvin temperature. A simple heat balance model assuming an independent phonon bath following the usual electron-phonon and Kapitza coupling laws can account for all experimental observations.
\end{abstract}
\date{\today}
\maketitle

The field of quantum nano-electronics is increasingly concerned with the question of controlling, manipulating and detecting thermal effects; electronic cooling using superconducting tunnel junctions \cite{NahumAPL94,MuhonenRPP12}, THz bolometers with an ultimate sensitivity \cite{WeiNatureNano08} and heat interferometers \cite{GiazottoNature12} being just three relevant examples. In every case, the question arises of the temperature of the phonon population coupled to the device electronic bath. At a temperature $T$, phonons have a thermal wavelength of the order of $hc/k_BT$, where $c$ is the material-dependent sound velocity. This dominant wavelength amounts to about 200 nm in Cu at 1 K, which is the order of magnitude of an usual device dimensions. It is thus generally assumed that phonons in a mesoscopic quantum device are strongly mixed with the substrate phonons and are thus thermalized at the bath temperature \cite{GiazottoRMP06}.

Phonon cooling is at the heart of the possibility of cooling a bulk detector \cite{ClarkAPL05} or a quantum device \cite{VercruyssenAPL11} supported on a membrane cooled by superconducting tunnel junctions. In this case, the substrate no longer acts as a heat sink and local phonons are actually cooled. This also holds for suspended metallic beams  \cite{KoppinenPRL09,MuhonenAPL09}. In similar electronic coolers but fabricated on a bulk substrate, a detailed thermal analysis of the cooling performance indicated that while electron-phonon coupling is the main bottleneck for the heat flow, a decoupling of the device phonons from the substrate is necessary to account for the data \cite{RajauriaPRL07}. A direct proof for phonon cooling was however lacking. More recently, measurements of the electron-phonon coupling strength in a thin metallic film at T $\approx$ 0.1-0.3 K demonstrated that it is nearly completely substrate-insensitive \cite{UnderwoodPRL11}. This finding adds support to the idea of an independent phonon population in the metallic thin film. While this phonon bath could exhibit specific properties due to its reduced dimensionality \cite{QuPRB05,HekkingPRB08}, only small deviations from bulk material properties were observed in suspended devices \cite{KarvonenPRL07}.

In this Rapid Communication, we demonstrate the existence of an independent phonon bath in a quantum device based on an electronic cooler. The device operation in both the cooling and the heating regimes enabled us to probe the thermal behavior over more than four decades of injected power. It is well described by bulk-like laws for electron-phonon coupling and Kapitza thermal resistance at every interface.

\begin{figure}[t]
\begin{center}
\includegraphics[width=0.93 \linewidth]{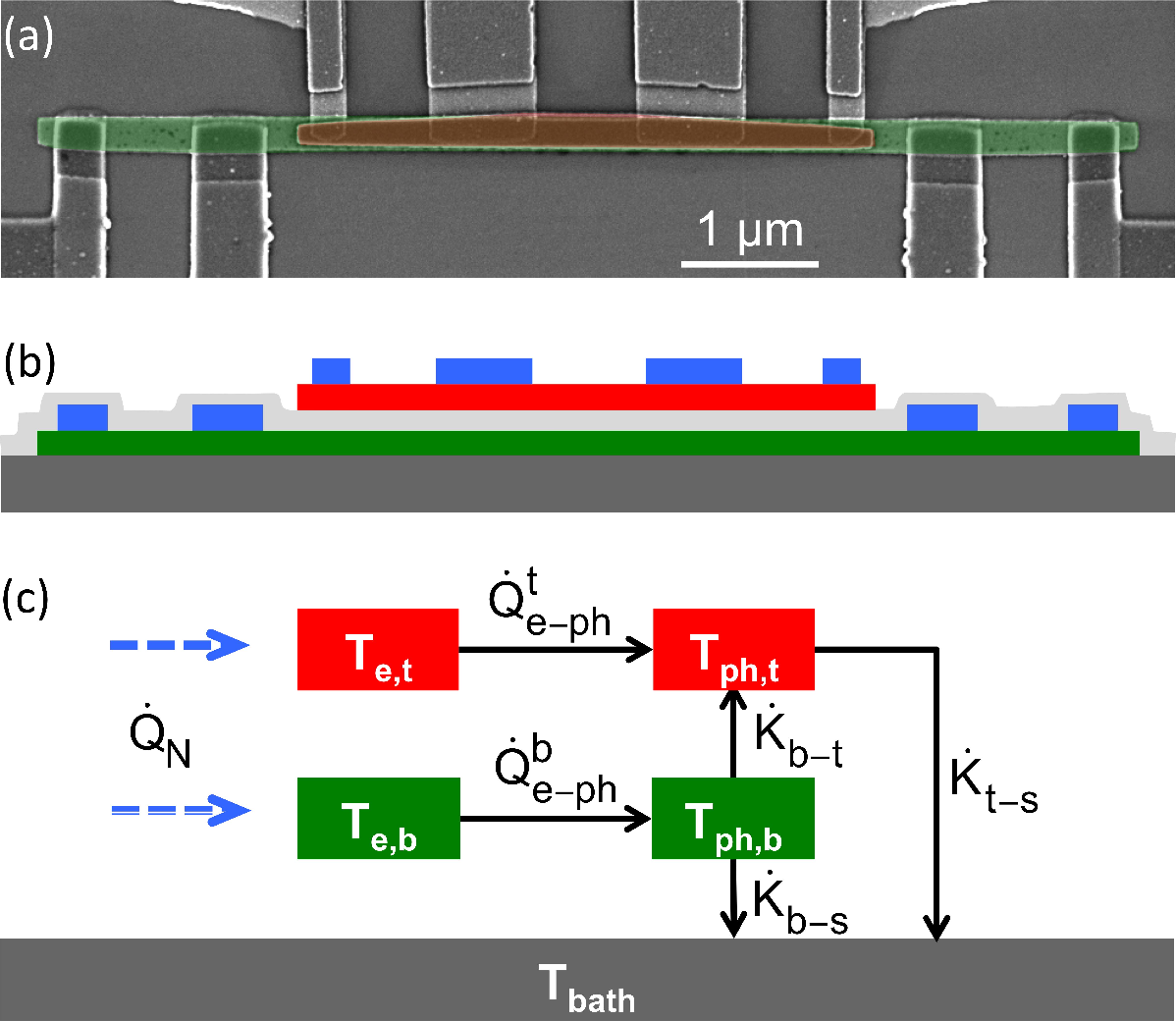}
\caption{(a) SEM image of the device. Each of the two normal islands (colorized in red or in green) is inserted between two sets of superconducting junctions. One junction pair is used as a cooler (or heater) and the other one is used as thermometer. (b) Schematic side-view of the set-up. The top (t) and bottom (b) levels are galvanically isolated from each other by a 40 nm thick layer of Si. (c) Heat transfer model. Electrons and phonons of the bottom (top) island at a respective temperature $T_{e,b}$ and $T_{ph,b}$ ($T_{e,t}$ and $T_{ph,t}$) exchange a heat power $\dot{Q}_{e-ph}^{b}$ ($\dot{Q}_{e-ph}^{t}$). Phonons of each island are coupled together via the Kapitza power $\dot{K}_{b-t}$, and to the bath phonons via $\dot{K}_{b-s}$ and $\dot{K}_{t-s}$.}
\label{fig:sample}
\end{center}
\end{figure}

Fig.~\ref{fig:sample} shows the device geometry made of two levels with a similar structure on a bulk Si substrate with 500 nm-thick SiO$_2$ oxide. At the bottom level, a Normal metal island (N) of dimensions 50 nm x 500 nm x 16 $\mu$m is connected to two pairs of Superconducting  electrodes (S) through tunnel junctions. The larger junctions pair is used for cooling or heating the metal. The other pair is used for electronic thermometry of the same metal \cite{NahumAPL93}. The top level has a similar geometry with the distinction of the metallic island being of smaller dimension: 50 nm x 400 nm x 8 $\mu$m. The device geometry was devised so that the two Cu islands are thermally coupled (only) through phonons. Each island is weakly coupled to the external world through tunnel junctions so that its electron population reaches a quasi-equilibrium state with a well-defined temperature. Moreover, one can heat or cool electrons in one island while monitoring its electronic temperature as well as the one of the other island.

The sample was fabricated through two successive electronic beam lithographies followed by a two-angle evaporation, with a precise alignment in-between. The two levels are separated by a 40 nm thick, e-beam evaporated layer of Si ensuring galvanic isolation. The junction pairs resistance is 0.74 and 2.35 $k\Omega$ for the bottom cooler and thermometer, 38.8 and 2.26 $k\Omega$ for the top ones respectively. Low temperature measurements were performed in a $^3$He cryostat with a base temperature below 260 mK. Filtering was provided by $\pi$-filters at room temperature and lossy micro-coaxial lines at the cold plate. Four-wire \mbox{d.c.} transport measurements were performed using home-made electronics combining three independent current bias sources, two of them being floating.

A N-I-S junction provides an easy way to perform electron thermometry in a normal metal \cite{NahumAPL93}. The charge current through such a junction biased at a voltage $V$ is: 
\begin{equation}                                                                                                                                                                                                                                                                                                                                                                                                                                                                                                                                                                                                                                   I=\frac{1}{eR_N}\int_{0}^\infty n_S(E,\Delta)[f_N(E-eV)-f_N(E+eV)]dE,
\label{eq:IV_theo}                                                                                                                                                                                                                                                                                                                                                                                                                                                                                                                                                                                                                                       \end{equation}
where $R_N$ is the normal state resistance, $f_N$ the electron energy distribution in the normal metal and $n_S(E,\Delta)$ the normalized BCS density of states in the superconductor. No Dynes parameter was taken into account. When using it as a thermometer, a N-I-S junction is biased at a small and constant current $I_{th}$. Here, the current was adjusted to 500 pA for the top and bottom thermometers so that the related heat flow (see below) can be safely neglected. The voltage drop $V_{th}$ in the thermometer junction pair is then measured and compared to its calibrated value against $T_{bath}$. This provides a measure of the electronic temperature $T_e$. Let us note that the calibration is realized close to equilibrium, with the superconductor being at the same temperature as the normal metal. In contrast, practical experiments are usually conducted in quasi-equilibrium conditions where, to a first approximation, only the normal metal temperature changes. If one considers temperatures above about half the superconductor critical temperature $T_c/2$, the related decrease of the superconductor gap $\Delta$ imposes to calculate the calibration voltage $V_{th}(T_e)$ using Eq.~(\ref{eq:IV_theo}). 

In addition, a N-I-S junction can be used to cool down or heat up a normal metal electron population. The heat current through a single junction writes:
\begin{equation}
\dot{Q}_N^0=\frac{1}{e^2R_N}\int_{-\infty}^\infty (E-eV)n_S(E, \Delta)[f_N(E-eV)-f_S(E)]dE,
\label{eq:QN_theo}             
\end{equation}
where $f_S$ is the energy distribution function in the superconductor \cite{MuhonenRPP12}. Eq.~\ref{eq:QN_theo} describes both the cooling and heating regimes. At large bias $V \gg \Delta/e$, the heat flow ${Q}_N^0$ is half the Joule power. As a whole, the full Joule power $IV$ is deposited in the device, so that a power $\dot{Q}_S=IV-\dot{Q}_N^0$ is transferred to the superconductor.

\begin{figure}[t]
\begin{center}
\includegraphics[width=\linewidth]{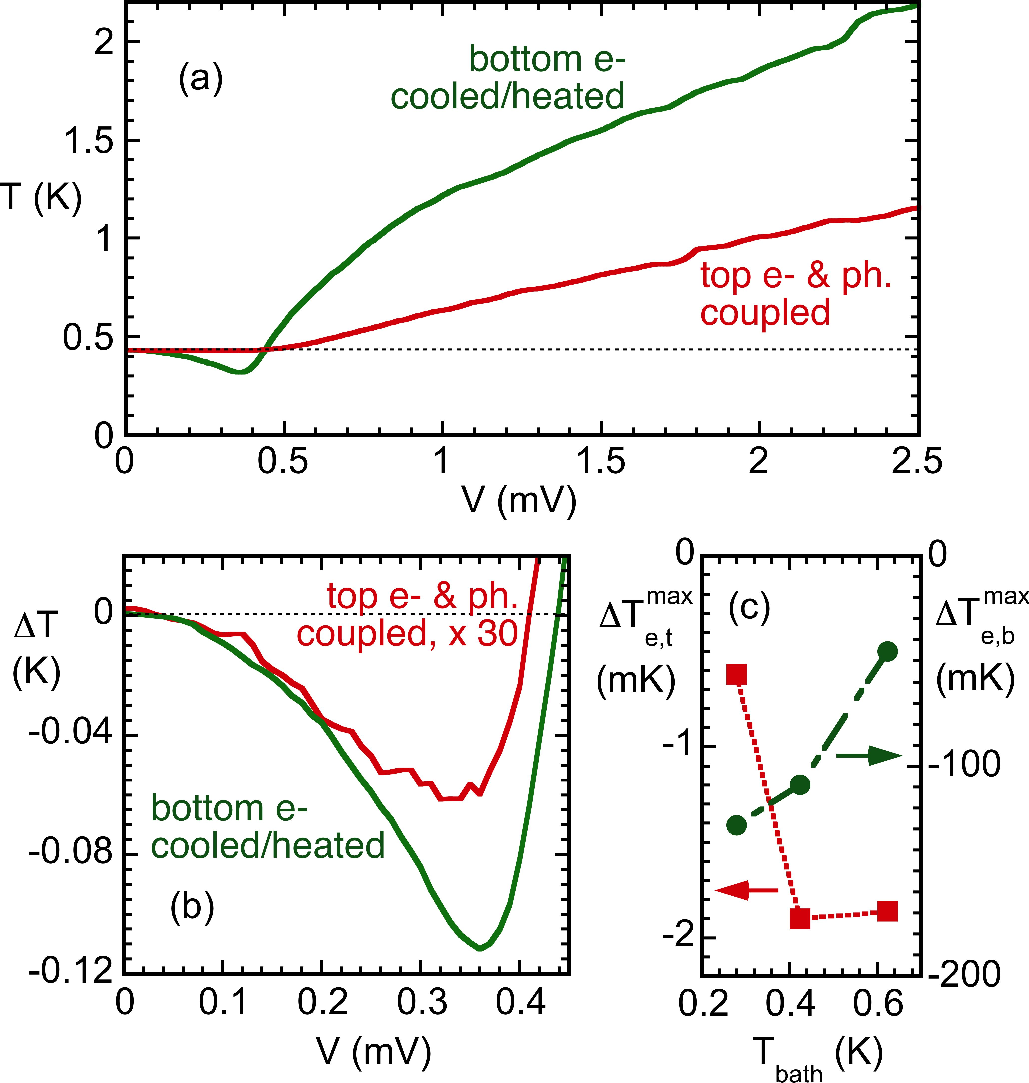}
\caption{(a) Electronic temperatures $T_{e,b}$ and $T_{e,t}=T_{ph,t}$ as a function of the voltage $V$ across the bottom cooler/heater junctions at a bath temperature $T_{bath}=432$ mK. (b) Temperature variations $\Delta T_{e,b}$ and $\Delta T_{e,t}=\Delta T_{ph,t}$ based on the same data, the latter temperature change being amplified by a factor 30. (c) Maximum of $\Delta T_{e,b}$ (circles) and $\Delta T_{e,t}$ (squares) as a function of the bath temperature.}
\label{T_vs_V}
\end{center}
\end{figure}

In the following, the indexes $e$ and $ph$ refer to electronic and phonon temperatures, while $s$, $b$ and $t$ refer to the substrate, to the bottom and top levels respectively. Our experiment consists in current-biasing one of the two level's cooler junction pair while monitoring simultaneously the related voltage drop $V$ as well as the two levels' electronic temperatures $T_{e,b}$ and $T_{e,t}$. As no power is directly injected in the unbiased electronic bath, its temperature is equal to the same metal's phonon temperature. In this way, our approach can be viewed as a phonon thermometry experiment. Fig.~\ref{T_vs_V}a shows the two sample levels' electronic temperatures $T_{e,b}$ and $T_{e,t}=T_{ph,t}$ as a function of the voltage drop $V$ applied to the bottom level, at a bath temperature $T_{bath}$=432 mK. For voltages below $2\Delta$, we observe the expected electronic cooling: the bottom electronic temperature $T_{e,b}$ goes well below the bath temperature $T_{bath}$, reaching a minimum of 320 mK. At voltages $V$ above $2\Delta$, we observe a hot-electron regime: the temperature $T_{e,b}$ increases and goes well above the bath temperature. In this regime, the top island electronic temperature $T_{e,t}$ also increases.

Remarkably, when the bottom level electronic temperature goes down, the top electronic temperature $T_{e,t}$ also diminishes with a variation $\Delta T_{e,t}$ reaching a maximum of - 2.0 mK, see Fig.~\ref{T_vs_V}b. This observation is the main experimental result of this paper. As the operation of the electronic cooler is dissipative as a whole, \mbox{i.e.} heat is dissipated in the chip, this observation cannot be related to an improper thermalization of the chip or of electrical leads \cite{SavinJAP06}. The observed cooling of the top level can only arise thanks to phonon cooling in the normal conductors of the device. This demonstrates the existence of a distinct phonon population in the mesoscopic metallic island of a quantum device. 

We have repeated this experiment at different bath temperatures. At higher bath temperatures $T_{bath}$, the maximum temperature decrease $\Delta T_{e,t}^{max}$ gets larger as opposed to $\Delta T_{e,b}^{max}$ that gets weaker, see Fig.~\ref{T_vs_V}c. As will be discussed later, this illustrates the fact that electron-phonon coupling and Kapitza thermal resistances have a different temperature dependence. In any case, the top level cooling remains of the order of a few mK, which is about 100 times smaller than the direct electronic cooling of the bottom level. We have also made the symmetric experiment by biasing the top island and monitoring its temperature as well as the temperature of the bottom metal. In this case, very little electron cooling could be observed because the top cooler junction pair was quite highly resistive \cite{PRB-Chaudhuri}. Phonon cooling could thus not be observed, but the heating regime was. In every configuration, we have checked that the measured differential conductance of the cooler/heater junctions compares well to Eq.~\ref{eq:IV_theo} prediction using the measured voltage-dependent normal metal temperature. 

\begin{figure}[t]
\begin{center}
\includegraphics[width=\linewidth]{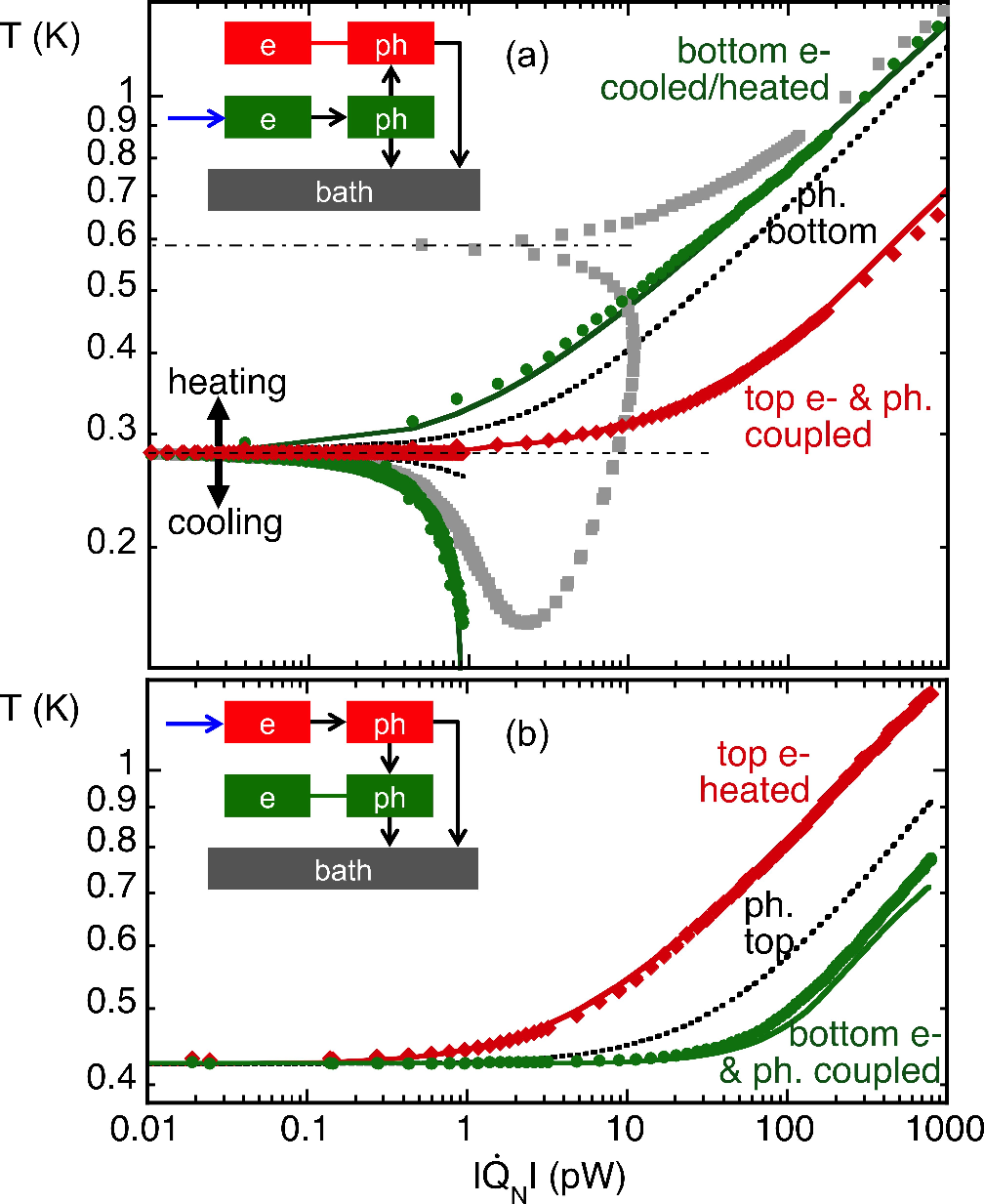}
\caption{(Color online) Electronic temperatures $T_{e,b}$ (green circles) and $T_{e,t}$ (red diamonds) as a function of the absolute value of the power injected in the bottom island (a) or in the top island (b). Grey squares in (a) show $T_{e,b}$ plotted as a function of the raw power $\dot{Q}_N^0$, while all other data are plotted as a function of the corrected power $\dot{Q}_N=\dot{Q}_N^0+\alpha Q_S$. In (a), the dash-dotted and the dotted lines corresponds to the point where the power $\dot{Q}_N^0$ or $\dot{Q}_N$ respectively change sign. The full lines are fits calculated using the thermal model discussed in the text and a single set of parameters. The black dotted lines indicate the calculated phonon temperature of the island that is cooled or heated. Bath temperature is 281 mK for (a) and 432 mK for (b). Insets depict the relevant experimental schemes.}
\label{T_vs_P}
\end{center}
\end{figure}

In order to analyze our data, let us plot the temperatures' evolutions with respect to the injected power. The power $\dot{Q}_N^0$ was calculated as a function of the coolers' bias by using Eq.~\ref{eq:QN_theo} and a value $\Delta=$ 214 $\mu eV$ for the superconducting gap, extracted from the individual junction characteristics. The measured electronic temperature in the biased level was taken as an input for the calculation. Fig.~\ref{T_vs_P} shows the evolution of every temperature measured when either the bottom level (a) or the top level (b) is biased. 

As a first approach, grey dots in Fig.~\ref{T_vs_P}a show the bottom electronic temperature plotted as a function of the absolute value of the raw power $\dot{Q}_N^0$ applied to the bottom level. This plot shows a striking behavior, with up to three values of temperature for a single power absolute value. The electronic temperature equals the bath temperature at a point where the calculated power is negative, \mbox{i.e.} where some cooling is expected. It is indeed expected that a fraction of the hot quasi-particules injected in the superconductor tunnels back in the normal metal, so that part of the related power is actually absorbed there. We have tried to describe this effect as a correction to the power that is proportional either to the current \cite{RajauriaPRB09} or to the power $\dot{Q}_S$ \cite{UllomPhysica00}. Only in the latter hypothesis does the electronic temperature plot as a function of the net power $\dot{Q}_N$ absolute value follow a single curve when one goes through the maximum cooling point, see Fig.~\ref{T_vs_P}a. In this case, the net power writes $\dot{Q}_N=\dot{Q}_N^0 +\alpha \dot{Q}_S$. The fit parameter value $\alpha$ = 0.087 is comparable to what appears in the literature \cite{UllomPhysica00}.  

In the following, we consider the thermal model depicted in Fig.~\ref{fig:sample}c. It is based on the assumption of two distinct phonon populations at quasi-equilibrium at temperatures $T_{ph,b}$ and $T_{ph,t}$ in the bottom and top metallic islands respectively. The heat extracted from or injected to the electronic baths is calculated as discussed above. No Andreev-current induced heat \cite{RajauriaPRL08} is taken into account as it shows up only below about 200 mK in this type of device. The direct photonic coupling \cite{PascalPRB11} between the two sample levels, with a maximum heat conductance of a single conductance quantum $k_B^2T\pi/6\hbar$= 0.28 pW/K at 300 mK, can also be neglected. One can also check that temperature drops within the substrate do not significantly contribute. Taking SiO$_2$ thermal conductivity to be $\kappa$ = 25 $T^\delta$ mW.m$^{-1}$.K$^{-1}$ with $T$ in Kelvin and $\delta$ = 1.91 \cite{Raychaudhuri}, the thermal conductance from the device contact area $a_{bs}$ of 8 $\mu$m$^2$ to the Si bulk substrate through the SiO$_2$ oxide thickness t = 500 nm is $G_{s} \approx \kappa a_{bs}/t$ = 40 nW.K$^{-1}$ at 0.3 K. At a typical 1 pW heat flow in the cooling regime, this corresponds to a negligible temperature drop of 0.025 mK. The bulk substrate contributes even less as, below 1 K, Si thermal conductivity is more than two orders of magnitude larger that the one of SiO$_2$.

In the model, we consider the usual laws for electron-phonon coupling and Kapitza resistance \cite{GiazottoRMP06}: we assume that electrons exchange with the phonons of the same metallic island a power $\dot{Q}_{e-ph}=\Sigma v[T_e^5-T_{p}^5]$, where $\Sigma$ is the electron-phonon coupling constant and $v$ is the island volume. We also assume that two neighbouring phonon populations (here x and y) are coupled through a Kapitza heat flow of the form $\dot{K}_{x-y}=k_{xy}a_{xy}[T_{ph,x}^4-T_{ph,y}^4]$, where $a_{xy}$ holds for the contact area between the two considered populations and $k_{xy}$ is an interface materials-dependent parameter. The top and bottom phonons are thus coupled together via the Kapitza coefficient $k_{bt}a_{bt}$, and to the substrate phonons via the coefficients $k_{ts}a_{ts}$ and $k_{bs}a_{bs}$, respectively. In the fit procedure, we have taken into account the device physical dimensions for calculating the volumes $v_{b}$, $v_{t}$ and the surfaces $a_{bt}$, $a_{bs}$, $a_{ts}$ of interest. The well-established value for the electron-phonon coupling constant in Cu $\Sigma$ =2 nW.$\mu$m$^{-3}$.K$^{-5}$ \cite{GiazottoRMP06} was used. We have chosen to take a common value for the substrate-bottom and bottom-top Kapitza parameters, so that the two fit parameters were $k_{bt}=k_{bs}$ and $k_{ts}$. Independent Kapitza parameters $k_{bt}$ and $k_{bs}$ could also be used, leading to the determination of different fit values, but with a low fit discrimination. The effect of phonon overheating in the superconducting electrodes was neglected. 

We obtained a very good fit of the whole data set over more than four orders of magnitude and the two signs for the net power, see continuous lines in Fig.~\ref{T_vs_P}a,b. The fit-derived Kapitza parameter $k_{bt}=k_{bs}$ = 45 pW.$\mu$m$^{-2}$.K$^{-4}$ describes the physical Si oxide-Cu and Cu-Si-Cu interfaces and compares well to values from the literature \cite{SwartzRMP89}. As for the top-substrate coupling, it describes the thermal leakage from the top island to the bath. As a whole, the coupling coefficient is $k_{ts}a_{ts}$ = 1100 pW.K$^{-4}$. Considering the contact area $a_{ts}$ to be the area of 1.2 $\mu$m$^2$ of the tunnel junctions connected to the top island, one obtains a Kapitza coefficient of 920 pW.$\mu$m$^{-2}$.K$^{-4}$, which is much larger than anticipated. Thus the heat transfer occurs presumably also along the continuous Si layer separating the two levels. This overall large thermal coupling is consistent with the modest amplitude of the phonon cooling observed in the top island, as compared to the corresponding cooling in the bottom island.

As for the experimental data measured at different bath temperatures, all data overlap in the high power regime and could be fitted with the same parameters set (not shown). In order to further test our hypotheses, we have also tried to replace the 5th-power temperature law for the electron-phonon coupling by a 6th or a 4th power \cite{SM}, which could respectively be justified in the cases of a strongly disordered metal \cite{SergeevPRB00} or of a direct electron-substrate phonon coupling \cite{SergeevPRB98}. No satisfying fit could be obtained under these assumptions. This was also the case when considering a single phonon population in the whole device or when neglecting the direct thermal coupling between the top island and the substrate.

From the thermal balance relations, one can calculate the phonon temperature variation in the cooled or heated metal, see dotted lines in Fig.~\ref{T_vs_P}a, b. The phonon population temperature decoupling is significant in the temperature range 0.3 - 1 K, consistently with previous estimates \cite{RajauriaPRL07}. At lower temperature, the electron-phonon bottle-neck makes the phonon temperature tend to the bath temperature. At a bath temperature below 100 mK, phonon cooling becomes negligeable.

In summary, we have devised an elaborate experiment providing access to the phonon temperature in a superconducting cooler device operated both in the cooling and in the heating regime. The experimental data demonstrate the existence of an independent phonon bath in the device. The thermal couplings are well described with the usual laws for electron-phonon coupling and Kapitza resistances. This new understanding can have significant outcomes in the analysis of quantum nano-electronic devices thermal behavior.

This work has received funding from the European Union Seventh Framework Programme under grant agreement INFERNOS, No. 308850, and through the low-temperature infrastructure MICROKELVIN. L. M. A. P. acknowledges a grant from R\'egion Rh\^one-Alpes. A. F. acknowledges a BQR grant from Grenoble INP. Samples were fabricated at Nanofab platform - CNRS. The authors thank B. Pannetier and M. Meschke for discussions, T. Fournier, T. Crozes, S. Dufresnes and J.-L. Mocellin for help in fabrication and measurement issues.

\end{document}